\documentclass[conference]{IEEEtran}
\IEEEoverridecommandlockouts
\usepackage{cite}
\usepackage{booktabs}
\usepackage{subcaption}
\usepackage{amsmath,amssymb,amsfonts}

\usepackage{graphicx}
\usepackage{textcomp}
\usepackage{xcolor}
\usepackage[most]{tcolorbox}

\usepackage{mathtools}
\usepackage{algorithm}

\usepackage{algpseudocode}
\usepackage{enumitem}
\usepackage{ragged2e}  

\usepackage{booktabs}
\usepackage{multirow}
\usepackage{dblfloatfix}
\usepackage{url}

 \usepackage{xcolor}

\def\BibTeX{{\rm B\kern-.05em{\sc i\kern-.025em b}\kern-.08em
    T\kern-.1667em\lower.7ex\hbox{E}\kern-.125emX}}

\definecolor{BaselineColor}{HTML}{ECECEC}   
\definecolor{DREColor}{HTML}{E7F0FF}        
\definecolor{COTColor}{HTML}{FFF0DA}        
\definecolor{MetaSynthColor}{HTML}{E6F4EA}  
\definecolor{CaseFrame}{HTML}{BDBDBD}

\newtcolorbox{casebox}[1]{
  enhanced, breakable,
  colback=white, colframe=CaseFrame,
  boxrule=0.6pt, arc=1.5pt,
  left=8pt, right=8pt, top=8pt, bottom=8pt,
  title={\textbf{#1}}
}

\newtcolorbox{legendbox}{
  colback=white, colframe=CaseFrame, arc=1.5pt, boxrule=0.5pt
}

\begin{document}

\title{MetaSynth: Multi-Agent Metadata Generation from Implicit Feedback in Black-Box Systems
}

\author{\IEEEauthorblockN{ Shreeranjani SrirangamSridharan*}
\IEEEauthorblockA{\textit{Shreeranjani.Srirang@walmart.com}\\
\textit{Walmart Global Tech}\\
Sunnyvale, California, USA\\
}
\and
\IEEEauthorblockN{ Ali Abavisani*}
\IEEEauthorblockA{\textit{ali.abavisani@walmart.com}\\
\textit{Walmart Global Tech}\\
Sunnyvale, California, USA\\
}
\and
\IEEEauthorblockN{Reza Yousefi Maragheh*}
\IEEEauthorblockA{\textit{Reza.YousefiMaragheh@walmart.com}\\
\textit{Walmart Global Tech}\\
Sunnyvale, California, USA\\
}
\and
\IEEEauthorblockN{Ramin Giahi}
\IEEEauthorblockA{\textit{Ramin.Giahi@walmart.com}\\
\textit{Walmart Global Tech}\\
Sunnyvale, California, USA\\
}
\and
\IEEEauthorblockN{Kai Zhao}
\IEEEauthorblockA{\textit{Kai.Zhao@walmart.com}\\
\textit{Walmart Global Tech}\\
Sunnyvale, California, USA\\
}
\and
\IEEEauthorblockN{Jason Cho}
\IEEEauthorblockA{\textit{Jason.Cho@walmart.com}\\
\textit{Walmart Global Tech}\\
Sunnyvale, California, USA\\
}
\and
\IEEEauthorblockN{Sushant Kumar}
\IEEEauthorblockA{\textit{Sushant.Kumar@walmart.com}\\
\textit{Walmart Global Tech}\\
Sunnyvale, California, USA\\
}
}

\maketitle

\begin{abstract}

Product titles and descriptions displayed in search results (also called meta title and descriptions) strongly shape engagement in search and recommendation platforms, yet optimizing them remains challenging. Ranking models are black boxes, explicit labels are unavailable, and feedback such as click-through rate (CTR) arrives only post-deployment. Existing template, LLM, and retrieval-augmented approaches either lack diversity, hallucinate attributes, or ignore whether candidate phrasing has historically succeeded in ranking. This leaves a gap in directly leveraging implicit signals from observable outcomes. We introduce MetaSynth, a multi-agent retrieval-augmented generation framework that learns from implicit search feedback. MetaSynth builds an exemplar library from top-ranked results, generates candidate snippets conditioned on both product content and exemplars, and iteratively refines outputs via evaluator–generator loops that enforce relevance, promotional strength, and compliance. On both proprietary e-commerce data and the Amazon Reviews corpus, MetaSynth outperforms strong baselines across NDCG, MRR, and rank metrics. Large-scale A/B tests further demonstrate +10.26\% CTR and +7.51\% clicks. Beyond metadata, this work contributes a general paradigm for optimizing content in black-box systems using implicit signals.

\end{abstract}


\section{Introduction}

Search and recommendation systems are central to online discovery, yet their internal ranking functions are typically opaque \cite{covington2016deep, doshi2017towards}. These systems disclose little about how items are ordered, but their outputs such as search engine result pages (SERPs) consistently reflect stable preferences in the way information is phrased and structured. Among the most impactful elements are meta titles and descriptions, the short snippets displayed to users at decision time, which strongly influence click behavior and downstream traffic\cite{clarke2007influence}. Optimizing these snippets represents a high-leverage intervention for organic acquisition. However, the optimization signal is fundamentally \emph{black-box}: ranking models expose no gradients, and observable metrics such as impressions or click-through rate (CTR) are only available post-deployment, where exploration is expensive and feedback is biased by confounding factors such as position and popularity \cite{joachims2017unbiased, islam2025correcting, zhao2024mitigate}. 

\begin{figure*}[t]
    \centering
    \includegraphics[width=0.9\linewidth]{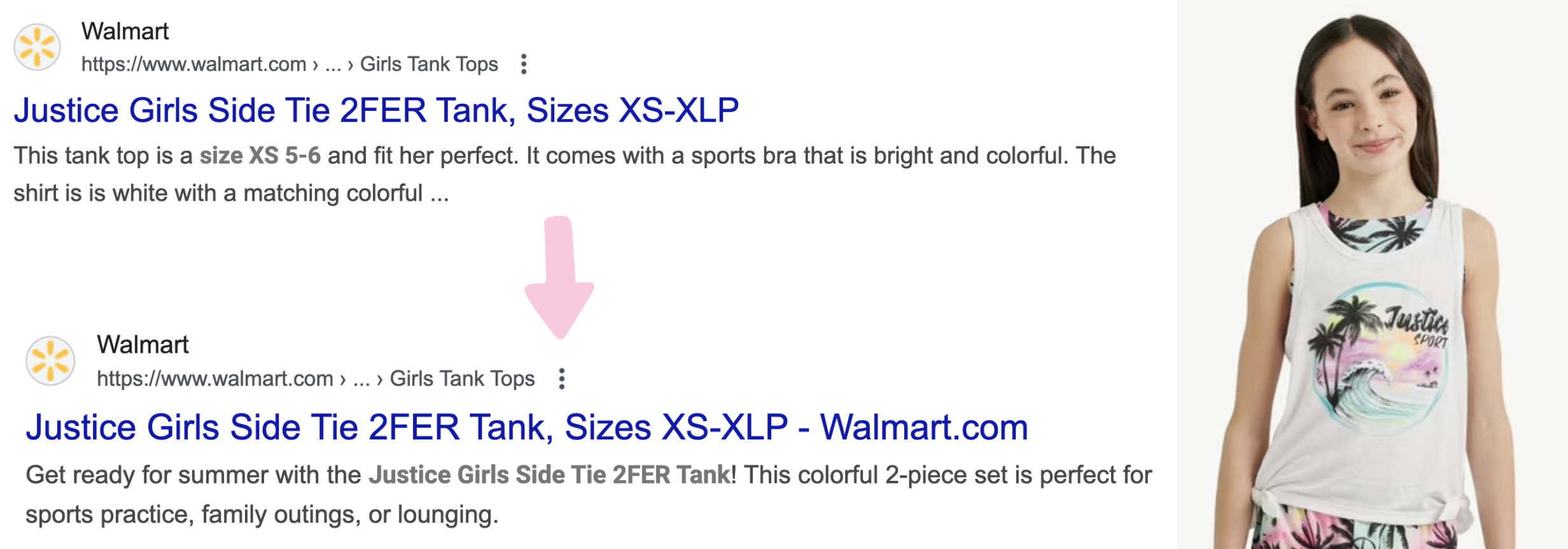}
    \caption{An example of how \textbf{MetaSynth} optimizes search engine meta descriptions. 
    The top snippet (pre-optimization) is factual but generic, while the bottom snippet (MetaSynth) 
    emphasizes promotional value, readability, and policy compliance. Such refinements directly impact 
    user engagement and search-driven traffic by producing coherent and persuasive messages.}
    \label{fig:fig1}
\end{figure*}

As illustrated in Fig.~\ref{fig:fig1}, even small stylistic changes to a snippet can significantly alter how users perceive and interact with results. In this example, the original snippet is accurate but lacks a strong promotional appeal, whereas the bottom snippet is a more engaging and policy-compliant description that highlights product attributes and use cases. This motivates the broader challenge: how can we systematically design models that learn from observable outcomes to optimize text for engagement, despite the black-box nature of modern rankers.

Existing strategies have notable limitations. Template-based generation provides consistency but lacks expressiveness and generalization across domains \cite{gatt2018survey, chen2021generate}. Prompt-only large language models (LLMs) generate fluent text but remain ungrounded, often hallucinating attributes or reverting to generic phrasing \cite{maragheh2025future}. Retrieval-augmented generation (RAG) improves factual grounding, but retrieval is usually based solely on content similarity, disregarding whether candidate styles have historically been rewarded in ranking outcomes \cite{yu2024evaluation, forouzandehmehr2025cal}. As a result, current methods fail to fully exploit the implicit supervision embedded in black-box outputs \cite{cao2025writing}.

We address this gap with \textbf{MetaSynth}, a multi-agent retrieval-augmented generation framework designed to ``play the black box search engine'' by systematically leveraging weak supervision from observable outcomes. MetaSynth constructs a ``exemplar library'' by harvesting (query, metadata) pairs from top-ranked results, thereby encoding implicit ranking preferences into a reusable corpus. For a new webpage and for constricting the search engine meta snippets of the webpage, an agentic retriever synthesizes plausible queries, retrieves the relevant examples from the library, and enriches the exemplar library when there is no relevant example in it. A constrained generator then produces candidate snippets conditioned on both product content and exemplar styles. These candidates are iteratively evaluated  by a panel of specialized evaluator agents that assess relevance, coverage, promotional tone, and compliance with brand constraints. Their feedback is fused by a consensus coordinator into targeted revisions, yielding an interpretable optimization loop guided by transparent  objectives.  

This framework reframes black-box optimization as a problem of ``learning from weak supervision'' while maintaining truthfulness and compliance. We evaluate MetaSynth through both offline experiments using proprietary and public data sets randomized online A/B tests. Offline studies enable principled ablation and iteration, while online deployment measures real-world impact on CTR and traffic.  

Empirical results show that MetaSynth consistently outperforms prompt-only LLMs, and standard RAG, achieving state-of-the-art performance across NDCG, MRR, and average rank. Online A/B tests further demonstrate statistically significant improvements of \textbf{+10.26\% CTR} and \textbf{+7.51\% clicks}, validating both effectiveness and scalability. Beyond metadata optimization, our work contributes a \emph{general paradigm for leveraging implicit preference signals in black-box systems}. We argue that this paradigm learning from weak but abundant observational cues opens new opportunities in ranking, recommendation, and personalization tasks where direct supervision is scarce but outcome-driven signals are observable.  

\noindent\textbf{Our main contributions are as follows:}
\begin{itemize}
    \item We propose \textbf{MetaSynth}, a novel multi-agent retrieval-augmented generation framework that exploits weak supervision from search and recommendation outcomes.  
    \item We introduce a \emph{exemplar success library} and autonomous retrieval strategy that encode implicit ranking preferences and dynamically expand coverage through agentic query generation, and provides weak supervision for the generation process. 
    \item We design an automated evaluator--generator refinement loop with consensus-driven feedback, ensuring that outputs satisfy relevance, fluency, and brand/policy guardrails.  
    \item We demonstrate the effectiveness of MetaSynth through both large-scale offline evaluations and online A/B tests, showing consistent improvements over strong baselines and significant real-world impact on CTR and clicks.  
\end{itemize}

\begin{figure*}[h]
    \centering
    \includegraphics[width=0.99\linewidth]{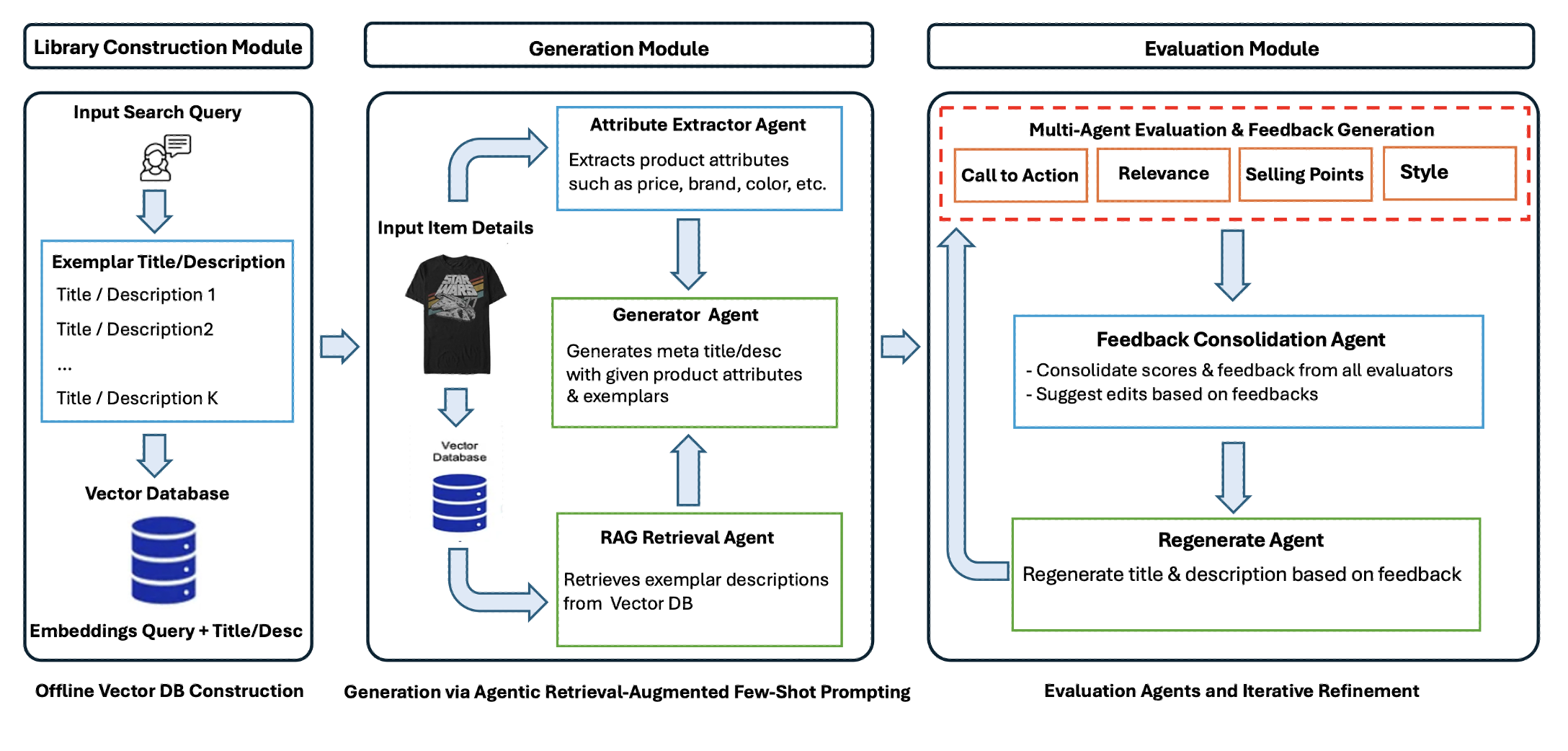}
    \caption{\small  MetaSynth Framework: Three main modules to generate and optimize meta titles and descriptions for items, according to seller's provided information, and constraints for better search engine ranking. Left block represents the RAG-based exemplar library creation. Middle block shows the generation module for both meta title and description based on seller provided data and exemplar library. Right block represents the evaluation/refinement loop, where multi evaluator agents evaluate the title/description for their assigned criteria, and if failed to satisfy that criteria, the feedback consolidation agent directs such title/description to re-generation agent along with the provided feedback from rejecting evaluator.}
    \label{fig:framework}
\end{figure*}

\section{Related Work}

Search and recommendation systems expose only their outputs, making it difficult to infer or optimize internal ranking preferences \cite{rolinek2020optimizing, sazanovich2021solving, giahi2025systems}. In this setting, user-facing snippets (meta titles and descriptions in SERPs) are known to shape attention and clicks, and thus downstream traffic \cite{chuklin2015anatomy, islam2019micro}. Prior work on query-biased summarization and snippet construction improves relevance and readability, but typically optimizes proxy text-quality metrics rather than outcome-driven objectives tied to ranking or clicks \cite{cao2016attsum, baumel2018query, chen2020abstractive}. Moreover, exploration in production is expensive and biased by position and popularity, motivating methods that can learn from observational data without full access to the ranker \cite{joachims2017unbiased, ermis2020learning}. 

Template-based NLG offers control and consistency for product and listing metadata but struggles to generalize stylistically across domains \cite{reiter1995nlg, deemter2005real, gajbhiye2021template}. Prompt-only LLMs increase fluency and diversity, yet can hallucinate attributes or regress to generic phrasing without grounding in historically successful styles \cite{chen2025carts, maragheh2025future}. Retrieval-augmented generation (RAG) improves faithfulness by conditioning on retrieved evidence, but standard retrieval is primarily similarity-driven and agnostic to whether candidate styles have performed well under ranking \cite{lewis2020retrieval, maragheh2025arag, forouzandehmehr2025cal}. 
Recent work examines leveraging observational signals for writing and recommendation \cite{loghmani2025aligning, cao2025writing, chen2023bias}, yet typically treats them as features for rerankers or classifiers rather than as priors for style-aware generation \cite{deldjoo2024recommendation, khan2023learning}.

Multi-agent LLM frameworks and self-refinement protocols use specialized roles (e.g., critics and verifiers) to improve reliability and adherence to constraints \cite{yu2025table, yuan2025reinforce, chen2025carts, wan2025mamm, giahi2025vl}. However, most evaluators optimize textual-quality proxies rather than outcome-aligned objectives and rarely close the loop with retrieval decisions \cite{motwani2024malt}. 

In contrast to mentioned work, MetaSynth treats the SERP itself as weak supervision about which phrasing and structure succeed under a black-box policy. It builds an \emph{exemplar library} from top-ranked results and conditions generation jointly on product content and \emph{outcome-informed} stylistic exemplars, thereby injecting an explicit, outcome-aligned prior into the generator rather than relying on content similarity alone. An agentic retriever synthesizes queries to retrieve and expand coverage when library support is sparse, while a constrained generator produces candidates that are subsequently refined by a panel of evaluator agents scoring relevance, coverage, promotional strength, and brand/policy compliance; a consensus coordinator converts this feedback into targeted revisions. This closes the loop between retrieval, generation, and evaluation without gradient access to the ranker or curated preference labels, yielding an interpretable and controllable optimization process that aligns style with observed preferences. As such, MetaSynth operationalizes weak supervision for style-aware generation and offers a practical pathway to optimize text for engagement in black-box ranking environments.

\section{Methodology}

In this section we introduce MetaSynth, a multi agent generation and evaluation framework that takes exemplar meta data as input to generate meta snippets along with an evaluator-refinement loop. Our approach contains three main components i) Library Construction Module ii) Generation Module and iii) Evaluation Module.  Fig~\ref{fig:framework} shows our entire framework design.

\subsection{Problem Definition}
Let $\mathcal{X}$ denote the set of retailer product pages for a target eCommerce platform. For a page $x \in \mathcal{X}$, with its associated textual meta data like product name, brand, category, seller description
$\mathbf{a}(x)$, the goal is to produce a meta-snippet
$y = (\tau, \delta)$ comprising a meta title $\tau$ and meta description $\delta$ that maximizes downstream organic acquisition
from a black-box search engine. We denote the (unknown) objective induced by the search engine and user behavior as
\begin{equation}
J(y \mid x) = \mathbb{E}[\mathrm{traffic} \mid x, y],
\end{equation}
which cannot be optimized directly because both ranking and user response are black-box. If the search engine was not a black box system, ideally we could have searched for optimal meta-snippet $y^*$ by maximizing $J(.)$:
\begin{equation*}
y* = \text{argmax}_y J(y \mid x)
\end{equation*}
We therefore construct a multi-agent
system that (i) learns from top-ranked search results as weak supervision for writing style, (ii) retrieves relevant exemplars, (iii)
generates candidate snippets, and (iv) iteratively evaluates and refines them under brand guardrails.

\begin{figure*}[h]
    \centering
    \includegraphics[width=0.75\linewidth]{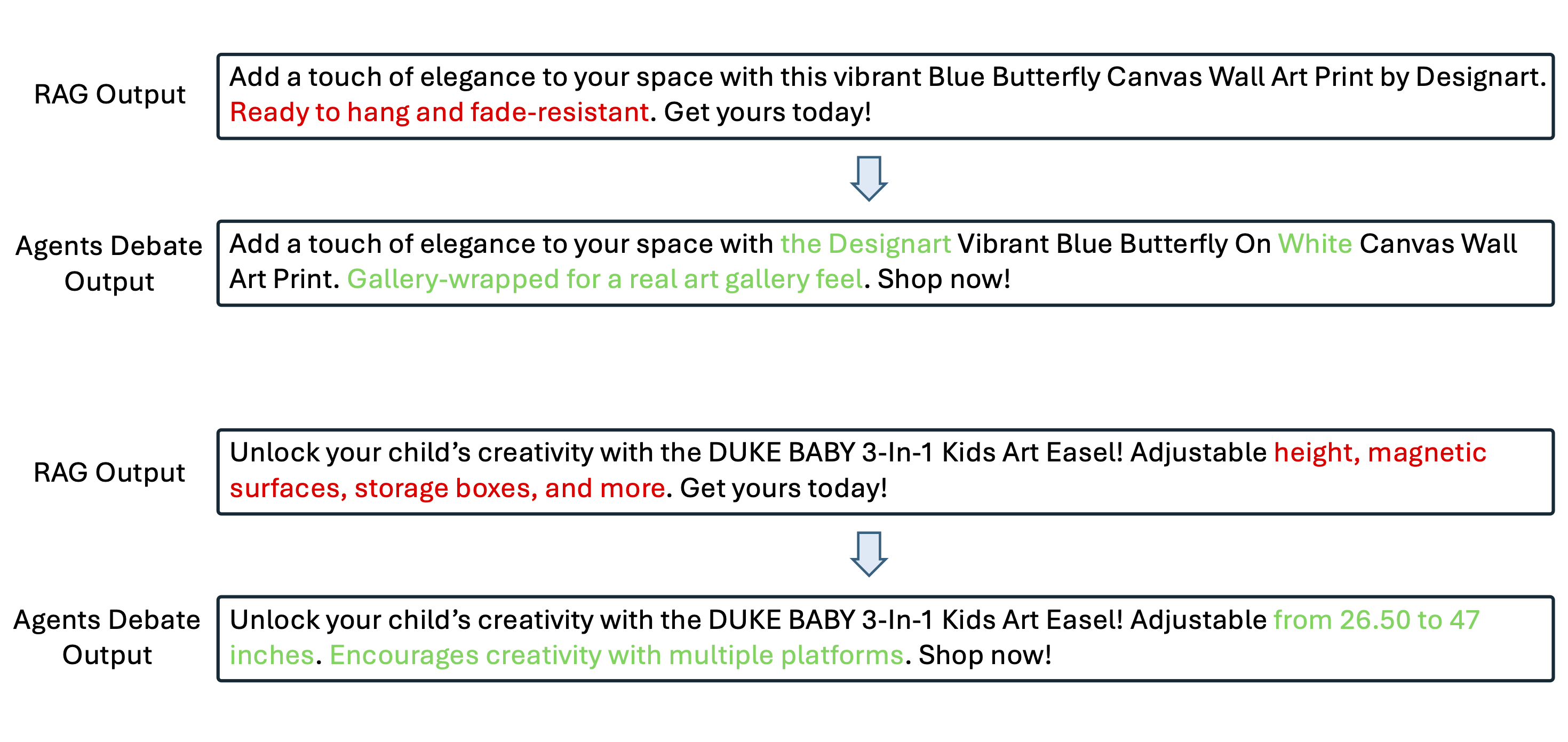}
    \caption{\small  Two examples showcasing the edits done by evaluation and refinement agents on RAG outputs, to make the description more accurate and engaging, according to seller's provided description.}
    \label{fig:examples}
\end{figure*}

Under this problem setting, one can model the search engine as a black-box function
\begin{equation}
\mathsf{Search}(q, K) \rightarrow \{(u_i, \tau_i, \delta_i, r_i)\}_{i=1}^{K},
\end{equation}
which, given a query $q$, returns the top-$K$ results with URL $u_i$, meta title $\tau_i$, meta description $\delta_i$, and rank $r_i=i$.
We maintain a library $\mathcal{L}$ of exemplars with entries
\(
e = (q, u, \tau, \delta, r).
\)

We embed all objects (including queries, product pages, and candidate snippets), we wish to compare into a shared vector space \(\mathbb{R}^d\). Formally, \(\mathbf{g}_q:\mathcal{Q}\!\to\!\mathbb{R}^d\) maps a textual query to a \(d\)-dimensional vector, \(\mathbf{g}_x:\mathcal{X}\!\to\!\mathbb{R}^d\) maps a product page (aggregating its structured and unstructured attributes), and \(\mathbf{g}_y:\mathcal{Y}\!\to\!\mathbb{R}^d\) maps a meta title--description pair. A common embedding space enables direct geometric comparisons across heterogeneous items: a query should lie near the products it meaningfully retrieves, and a high-quality snippet should lie near exemplars that reflect the desired style and content. While we do not assume a specific embedding model, we require that these maps are calibrated so that proximity corresponds to semantic relatedness. We measure proximity using cosine similarity.

Brand guardrails are denoted by set $\mathcal{B}$. In practice, these guardrails are stated by constraints, requirements, and acceptance thresholds denoted by  $\mathcal{H}, \mathcal{R}$, and $\boldsymbol{\alpha}$ respectively . The set \(\mathcal{H}\) captures \emph{hard} prohibitions (e.g., legally sensitive claims, banned phrases), any of which yields immediate rejection. The set \(\mathcal{R}\) specifies \emph{required} elements (e.g., presence of a call to action, inclusion of brand name) that must be verifiably present in the snippet. The vector \(\boldsymbol{\alpha}\) contains per-criterion thresholds for continuous quality scores (e.g., minimum relevance to the page, minimum promotional strength, minimum style compliance). During evaluation, the system computes a score vector and checks it against \(\boldsymbol{\alpha}\) while enforcing \(\mathcal{H}\) and \(\mathcal{R}\); only candidates satisfying all hard/required constraints and meeting or exceeding the thresholds are accepted, ensuring optimization for search performance remains aligned with brand and policy requirements.

\subsection{Library Construction Module: Offline Vector DB Construction} \label{LCM}

Given a seed set of popular queries $\mathcal{Q}_{\mathrm{S}}$, the Library Construction agent issues calls to the black-box engine and builds
\(
\mathcal{L}=\bigcup_{q\in\mathcal{Q}_{\mathrm{S}}}\,
\{(q, u_i, \tau_i, \delta_i, r_i)\}_{i=1}^{K_{\mathrm{lib}}}.
\)
For saving each exemplar $i$, both its meta title $\tau_i$ and meta   description $\delta$ are concatenated (denoted by $\oplus$) and embedded using  $\mathbf{g}_{y
_i}(e_i) = \mathbf{g}_{y_i}(\tau_i\!\oplus\!\delta_i)$. We do not save the embedding vector if the cosine similarity with the most similar item in the library is more than a threshold $\epsilon_{\mathrm{dup}}$. In other words, $e_i$ is considered a duplicate of existing $e_j$ in the library if :
\(
\mathrm{sim}(\mathbf{g}_y(e_i),\mathbf{g}_y(e_j))>\epsilon_{\mathrm{dup}}
\). 

We also index a query-to-exemplar map
\(
\mathcal{I}(q)=\{e\in\mathcal{L}: e.\!q=q\}
\)
and a global ANN index over $\mathbf{g}_q(q)$ to support fast retrieval.

\subsection{Generation Module: Generation via Agentic Retrieval-Augmented Few-Shot Prompting}\label{GMA}

\subsubsection{Target Query Detection and Library Construction} To select a query for a target product page $x$, first its textual meta-data embedding vector is computed $\mathbf{z}_x=\mathbf{g}_x(a(x))$. Then, the most similar query $q^*$ to $\mathbf{z}_x$ is obtained and the similarity score between $q^*$ and $x$ is stored. In other words, we have
\begin{equation}
    \begin{cases}
        q^\star &\;=\; \arg\max_{q\in \mathrm{dom}(\mathcal{I})}\;\mathrm{sim}\!\big(\mathbf{z}_x, \mathbf{g}_q(q)\big),\\
        s^\star & \;=\; \max_{q}\;\mathrm{sim}\!\big(\mathbf{z}_x, \mathbf{g}_q(q)\big).
    \end{cases}
\end{equation}

Given a target similarity threshold $\tau_q \in\!(0,1)$, if $s^\star < \tau_q$ (i.e., no sufficiently similar query exists),
the generation agent constructs candidate queries from attributes, by using $\mathrm{Expand}$ prompt template that generates new relevant queries $q_{\mathrm{new},x}$ based on the product page's associated textual data $a(x)$. In other words, 
\begin{equation}
    q_{\mathrm{new},x} \;=\;\mathrm{Expand}\!\big(\mathbf{a}(x)\big),
\end{equation}
and then invokes $\mathsf{Search}$ for each $q_{\mathrm{new},x}$. This will ensure exemplar writings for all rounds of writing.

For each $q\in q_{\mathrm{new},x}$, we augment $\mathcal{L}$ with the top-$K_{\mathrm{aug}}$ results extracted by $\mathsf{Search}(q, K)$ call and update the ANN index. 

For the case where there exists a similar query in the library ($s^\star\ge \tau_q$), we instead take the most relevant query $q^*$ itself as well as all other similar queries passing the threshold $\tau_q$. Then, we construct the Exemplar set as described below using obtained set. 

\subsubsection{Exemplar set construction}
Let
\(
\mathcal{F}_x
\)
be the candidate exemplars set for webpage $x$. The $m$ few-shot exemplars selected by a greedy algorithm using Maximal Marginal Relevance (MMR) \cite{carbonell1998use} to balance relevance and diversity:
\begin{equation}\label{howtosel}
    \begin{split}
        \mathrm{MMR}(e \mid \mathcal{F}) \;=\; & \lambda\cdot \mathrm{sim}  \!\big(\mathbf{z}_x,\mathbf{g}_y(e)\big)\\
        \;& -\; (1-\lambda)\cdot \max_{e'\in\mathcal{F}} \mathrm{sim}\!\big(\mathbf{g}_y(e),\mathbf{g}_y(e')\big),
    \end{split}
\end{equation}
where $\mathcal{F}$ is the selected growing set and $\lambda\in[0,1]$. We iterate $m$ steps to obtain $\mathcal{F}_x$. This will ensure that we have a diverse enough set of successful examples to be passed to meta-snippet generation. Note that when selecting the exemplar set, we can potentially inflate the similarity score ($\lambda\cdot \mathrm{sim}  \!\big(\mathbf{z}_x,\mathbf{g}_y(e)\big)$) as a function of ranking, $r$, of the exemplar writing recorded in the exemplar library. In this way, we also prioritize writings with better rankings. One can potentially deflate too in cases where popularity bias is severely inherent in search engine's recommendations.  

\begin{algorithm}
\caption{Concise Multi-Agent Meta-Snippet Generation}
\label{alg:concise}
\begin{algorithmic}[1]
\Require Seed queries $\mathcal{Q}_{\mathrm{S}}$; cutoffs $K_{\mathrm{lib}},K_{\mathrm{hit}},K_{\mathrm{aug}}$; thresholds $\epsilon_{\mathrm{dup}},\tau_q$; few-shot size $m$; MMR weight $\lambda$; max iters $K_{\max}$; guardrails $\mathcal{B}=(\mathcal{H},\mathcal{R},\boldsymbol{\alpha})$
\Statex \textbf{Offline: Library}
\For{$q \in \mathcal{Q}_{\mathrm{S}}$}
  \State $R \leftarrow \mathsf{Search}(q,K_{\mathrm{lib}})$
  \For{$(u,\tau,\delta,r)\in R$}
     \If{$\max_{e\in\mathcal{L}}\mathrm{sim}\!\big(\mathbf{g}_y(\tau\!\oplus\!\delta),\mathbf{g}_y(e)\big) < \epsilon_{\mathrm{dup}}$}
        \State add $e{=}(q,u,\tau,\delta,r)$ to $\mathcal{L}$; \ $\mathcal{I}(q)\!\leftarrow\!\mathcal{I}(q)\cup\{e\}$
     \EndIf
  \EndFor
\EndFor
\Statex \textbf{Online: Page $x$ (URL $u_x$)}
\State $\mathbf{z}_x \!\leftarrow\! \mathbf{g}_x(\mathbf{a}(x))$;\quad $s^\star \!\leftarrow\! \max_{q}\mathrm{sim}(\mathbf{z}_x,\mathbf{g}_q(q))$
\If{$s^\star < \tau_q$} \Comment{no similar repo query $\Rightarrow$ agentic search}
   \State $\mathcal{Q}_{\mathrm{new}} \!\leftarrow\! \mathrm{Expand}(\mathbf{a}(x))$
   \State $\mathcal{Q}_{\mathrm{S}}(x) \!\leftarrow\! \{\, q\!\in\!\mathcal{Q}_{\mathrm{new}} : u_x \in \text{top-}K_{\mathrm{hit}}\text{ of }\mathsf{Search}(q,K_{\mathrm{hit}})\,\}$
   \State augment $\mathcal{L}$ with top-$K_{\mathrm{aug}}$ items (excluding $u_x$) from those searches; update $\mathcal{I}(\cdot)$
\Else
   \State $\mathcal{Q}_{\mathrm{S}}(x) \!\leftarrow\! \{\, q : \mathrm{sim}(\mathbf{z}_x,\mathbf{g}_q(q)) \ge \tau_q \,\}$
\EndIf
\State $\mathcal{E}(x) \!\leftarrow\! \bigcup_{q\in \mathcal{Q}_{\mathrm{S}}(x)} \mathcal{I}(q)$;
\State $\mathcal{F}_x \!\leftarrow\! \mathrm{MMR\_Select}(\mathcal{E}(x),\mathbf{z}_x,m,\lambda)$
\State $y^{(0)} \!\leftarrow\! G(x,\mathcal{F}_x,\mathcal{B})$
\For{$t=0$ \textbf{to} $K_{\max}-1$}
   \State $(\mathbf{s}^{(t)},\mathbf{c}^{(t)}) \!\leftarrow\! E_{\phi}(y^{(t)},x,\mathcal{B})$
   \If{$\big(\forall k,\ s_k^{(t)} \!\ge\! \alpha_k\big)$}
      \State \textbf{break} \Comment{accepted}
   \Else
      \State $y^{(t+1)} \!\leftarrow\! G(x,\mathcal{F}_x,\mathcal{B},y^{(t)},\mathbf{c}^{(t)})$
   \EndIf
\EndFor
\State \Return $y^{(t)}$
\end{algorithmic}
\end{algorithm}

Let $G$ be a generation function conditioned on (i) product content, (ii) $\mathcal{F}_x$, and (iii) brand guardrails $\mathcal{B}$.
The initial meta-snippet is obtained by  
\begin{equation}
y^{(0)} \;=\; G\big(x, \mathcal{F}_x, \mathcal{B}\big),
\end{equation}
where the prompt includes structured slotting of the associated text with webpage $x$, $a(x)$, and the selected exemplars $\mathcal{F}_x$, and guardrails $\mathcal{B}$.

Please note that The relevance filter requiring $u_x$ to appear in the top-$K_{\mathrm{hit}}$ enforces that only queries demonstrably leading to the target page are retained; the few-shot pool $\mathcal{F}_x$ is then assembled from the \emph{top results of those queries}, serving as weakly supervised exemplars of effective writing styles.

\subsection{Evaluation Module: Evaluation Agents and Iterative Refinement}

The Evaluation agent $E_{\phi}$ scores a candidate $y$ for page $x$ along $K$ criteria and returns a score vector
\(
\mathbf{s}(y,x)\in[0,1]^K
\)
and textual feedback $\mathbf{c}(y,x)$.
We instantiate the following four primary criteria: (i) $s_{\mathrm{rel}}(y,x) \in [0,1]$ which evaluates if the generated meta-snippet, $y$, is  relevant to the target item page, $x$, (ii) $s_{\mathrm{promo}}(y) \in [0,1]$  which evaluates if $y$ has a promotional tone, (iii) $s_{\mathrm{cta}}(y) \in \{0,1\}$ which evaluates if $y$ has a call-to-action (phrases like ``buy now''), and (iv)  $s_{\mathrm{brand}}(y;\mathcal{H}) \in [0,1]$ which evaluates if $y$ is abiding the brand/style guidelines.

The generated feedback at each round $t$ by evaluators is stored $\mathbf{c}^{(t)}=\mathbf{c}(y^{(t)},x)$ and given this feedback the generator produces a revised meta-snippet:
\begin{equation}
y^{(t+1)} \;=\; G\!\big(x, \mathcal{F}_x, \mathcal{B}, y^{(t)}, \mathbf{c}^{(t)}\big).
\end{equation}

In this case, $\mathbf{c}^{(t)}$ is injected as structured constraints (e.g., “increase promotional strength,” “insert CTA,” “remove forbidden term $h$”).
The cycle stops at iteration $t^\star$ if either (i) if the Evaluator accepts the generated text on all criteria or (ii) iteration hits the max iteration budget $K_{\max}$. Optionally, a stagnation rule halts the iterations if enough improvement  does not happen at all (or enough) for two consecutive steps.
Fig. \ref{fig:examples} shows an example of how evaluation module modifies the initial generated output of Section III C.

\subsection{Generalization to other Problem Settings}
The proposed MetaSynth framework can demonstrate strong generalization beyond search engine optimization tasks. At its core, the methodology addresses a fundamental challenge in digital optimization: evaluating content quality when the only available signal is implicit feedback from a black-box system. This paradigm extends naturally to numerous domains where explicit quality labels are absent but performance metrics exist. For instance, in social media marketing, engagement metrics such as likes, shares, and comments serve as implicit feedback signals that can be leveraged to construct exemplar libraries of high-performing posts, enabling automated evaluation of new content against proven successful patterns. Similarly, in digital advertising, click-through rates and conversion metrics provide implicit feedback that can guide the creation of exemplar-based evaluation systems for ad copy, creative assets, and landing page content. E-commerce product listings, email marketing campaigns, and even content recommendation systems all share this characteristic of having measurable outcomes without explicit quality annotations. The RAG-based approach of retrieving relevant high-performing examples and evaluating against domain-specific criteria (relevance, uniqueness, key selling points, call-to-action presence) represents a generalizable framework applicable wherever implicit feedback signals can proxy for content effectiveness, making it a versatile solution for optimization challenges across the digital marketing ecosystem.

\begin{table*}[t]
\centering
\caption{Offline evaluation results: LLM-as-a-judge metrics across Amazon and proprietary datasets for meta titles and meta descriptions.}
\label{tab:combined_evaluation_results}
\begin{tabular}{@{}llcccccc@{}}
\toprule
\multirow{2}{*}{\textbf{Dataset}} & \multirow{2}{*}{\textbf{Approach}} & \multicolumn{3}{c}{\textbf{Meta Title}} & \multicolumn{3}{c}{\textbf{Meta Description}} \\
\cmidrule(lr){3-5} \cmidrule(lr){6-8}
& & \textbf{NDCG} & \textbf{MRR} & \textbf{Avg Rank} & \textbf{NDCG} & \textbf{MRR} & \textbf{Avg Rank} \\
\midrule
\multirow{4}{*}{\textbf{Amazon Fashion}} & Baseline & 0.5970 & 0.4617 & 2.5284 & 0.5504 & 0.4013 & 2.8355 \\
& DRE & 0.5095 & 0.3505 & 3.2731 & 0.5673 & 0.4261 & 2.9335 \\
& COT & 0.6280 & 0.5046 & 2.4911 & 0.6443 & 0.5282 & 2.5221 \\
& MetaSynth & \textbf{0.8190} & \textbf{0.7601} & \textbf{1.7025} & \textbf{0.7911} & \textbf{0.7213} & \textbf{1.6987} \\
\midrule
\multirow{4}{*}{\textbf{Amazon Toys}} & Baseline & 0.6302 & 0.5051 & 2.3450 & 0.5873 & 0.4492 & 2.5834 \\
& DRE & 0.5174 & 0.3606 & 3.2197 & 0.5972 & 0.4657 & 2.7714 \\
& COT & 0.5911 & 0.4562 & 2.7260 & 0.5804 & 0.4450 & 2.9459 \\
& MetaSynth & \textbf{0.8149} & \textbf{0.7551} & \textbf{1.7015} & \textbf{0.7881} & \textbf{0.7169} & \textbf{1.6921} \\
\midrule
\multirow{4}{*}{\textbf{Amazon Home}} & Baseline & 0.6648 & 0.5510 & 2.2187 & 0.5609 & 0.4147 & 2.7284 \\
& DRE & 0.5486 & 0.4005 & 3.0045 & 0.5885 & 0.4543 & 2.8140 \\
& COT & 0.5846 & 0.4478 & 2.8052 & 0.6048 & 0.4764 & 2.7941 \\
& MetaSynth & \textbf{0.7598} & \textbf{0.6809} & \textbf{1.9694} & \textbf{0.7996} & \textbf{0.7316} & \textbf{1.6627} \\
\midrule
\multirow{4}{*}{\textbf{Proprietary}} & Baseline & 0.5762 & 0.4352 & 2.6834 & 0.4771 & 0.3160 & 3.2655 \\
& DRE & 0.5553 & 0.4102 & 2.9622 & 0.5066 & 0.3552 & 3.1309 \\
& COT & 0.6527 & 0.5381 & 2.4191 & 0.7117 & 0.6243 & 1.9640 \\

& MetaSynth & \textbf{0.7631} & \textbf{0.6882} & \textbf{1.9716} & \textbf{0.7835} & \textbf{0.7204} & \textbf{1.6416} \\
\bottomrule
\end{tabular}
\label{Table:Evaluation}
\end{table*}

\section{Experiments and Results}
\subsection{Data and Experiment Setting}

We conduct experiments on two datasets to comprehensively evaluate MetaSynth.  

\begin{itemize}
    \item \textbf{Proprietary dataset:} A large-scale e-commerce catalog containing 40{,}000 items. We sample an equal proportion of products from four diverse categories---Clothing, Electronics, Toys, and Home \& Garden. Each item is associated with rich metadata, including product titles, specifications, brand information, and customer reviews.  
    \item \textbf{Amazon Review dataset \cite{hou2024bridging}:} A widely used public benchmark curated by McAuley Lab. We extract 30,000 items across three domains, Home, Toys, and Fashion, with 10,000 items in each category. This dataset provides structured metadata such as item descriptions, prices, and review text, making it complementary to our proprietary corpus.  
\end{itemize}

For preprocessing, we apply standard NLP techniques to remove non-ASCII characters, special symbols, and noise. For each item, we employ an LLM to generate the most likely user query that could lead to that item. We then identify the closest matching query from our VectorDB (as described in section \ref{LCM}) and retrieve the top-$k$ exemplar meta titles and descriptions. These exemplars serve as weakly supervised demonstrations for the \textbf{Generation Module} (section \ref{GMA}), ensuring that outputs are conditioned on successful historical styles. This setup allows us to test MetaSynth on both controlled proprietary environments and an open, publicly reproducible benchmark.

\subsection{Evaluation Metrics}

We compare MetaSynth against three representative baselines:  

\begin{enumerate}
    \item \textbf{Vanilla (Baseline):} A single LLM call that generates meta titles and descriptions directly from item metadata, without retrieval or reasoning.  
    \item \textbf{DRE (Direct Retrieval Enhancement) \cite{gao2024dre}:} An approach that extracts keywords from items and incorporates them into generation prompts, thereby enriching outputs with content-derived terms.  
    \item \textbf{CoT (Chain-of-Thought prompting) \cite{wei2022chain}}: A reasoning-based method that guides LLMs through structured intermediate steps, improving factual alignment between metadata and generated snippets.  
\end{enumerate}

To ensure consistent evaluation, we use \textbf{GPT-4.1-mini} as a judgment model, applying the LLM-as-a-judge paradigm \cite{zheng2024mtbench} to rank generated outputs. We report three widely accepted ranking metrics:

\begin{itemize}
    \item \textbf{NDCG (Normalized Discounted Cumulative Gain):} Measures the overall quality of ranked outputs, giving higher weight to relevant outputs that appear at the top of the list. Scores are normalized between 0 and 1, with higher values indicating stronger alignment with ideal rankings.  
    \item \textbf{MRR (Mean Reciprocal Rank):} Captures how quickly the first highly relevant output appears in the ranking. An MRR close to 1 implies that the correct snippet is almost always ranked first.  
    \item \textbf{Average Rank:} Records the average position of generated outputs across all items. Lower values indicate better performance, as strong methods consistently place outputs near the top.  
\end{itemize}

Together, these metrics provide complementary perspectives: NDCG highlights ranking quality, MRR emphasizes efficiency in surfacing relevant snippets, and Average Rank evaluates overall placement robustness.

\subsection{Experiment Results}

On the Amazon Review dataset (See Table \ref{Table:Evaluation}) , Baseline establishes a modest benchmark (e.g., NDCG 0.5970, MRR 0.4617 on Fashion titles), but its lack of explicit reasoning often results in suboptimal ranking. DRE provides incremental gains in some settings, such as improved meta description retrieval on Home (NDCG 0.5885 vs. 0.5609 for Baseline), but overall lags behind other methods, as its heuristic adjustments fail to capture deeper semantic structure. COT delivers more consistent improvements by incorporating structured reasoning, achieving higher retrieval quality (e.g., Fashion meta description MRR of 0.5282 vs. 0.4013 for Baseline). However, while COT narrows the gap, its reasoning alone cannot fully address domain-specific nuances in product data.

MetaSynth achieves the strongest results across all datasets, significantly outperforming previous methods. On Fashion titles, MetaSynth reaches NDCG of 0.8190 and MRR of 0.7601, compared to 0.6280 and 0.5046 for COT. A similar trend holds for meta descriptions, where MetaSynth (NDCG 0.7911, MRR 0.7213) consistently outperforms all baselines. The improvements extend to Toys (title MRR 0.7551 vs. 0.4562 for COT) and Home (description NDCG 0.7996 vs. 0.6048 for COT), with MetaSynth reducing average rank to ~1.7 across domains. These results highlight MetaSynth’s ability to generalize across categories while preserving fine-grained detail in retrieval.

On the e-commerce Proprietary dataset, the differences become even clearer. On this dataset, Vanilla and DRE show relatively weak performance, with DRE only marginally improving over Vanilla in description generation. COT achieves notable gains, particularly in meta descriptions (NDCG of 0.7117 and MRR of 0.6243), demonstrating the utility of structured reasoning in this domain. However, MetaSynth delivers the strongest results across all metrics, achieving the highest NDCG (0.7631 for titles, 0.7835 for descriptions) and the lowest average rank (1.9716 for titles, 1.6416 for descriptions). Still, MetaSynth delivers the largest gains, achieving  the highest NDCG and MRR. 

The offline evaluation metrics clearly show that our proposed approach MetaSynth consistently performs in all of them beating all benchmark models indicating the importance of exemplar library and evaluator feedback loop.  Although benchmarks like COT and DRE fare better than Vanilla,  MetaSynth yields state-of-the-art retrieval performance across all datasets and metrics.

\begin{figure*}[ht!]
    \centering
    \includegraphics[width=0.99\linewidth]{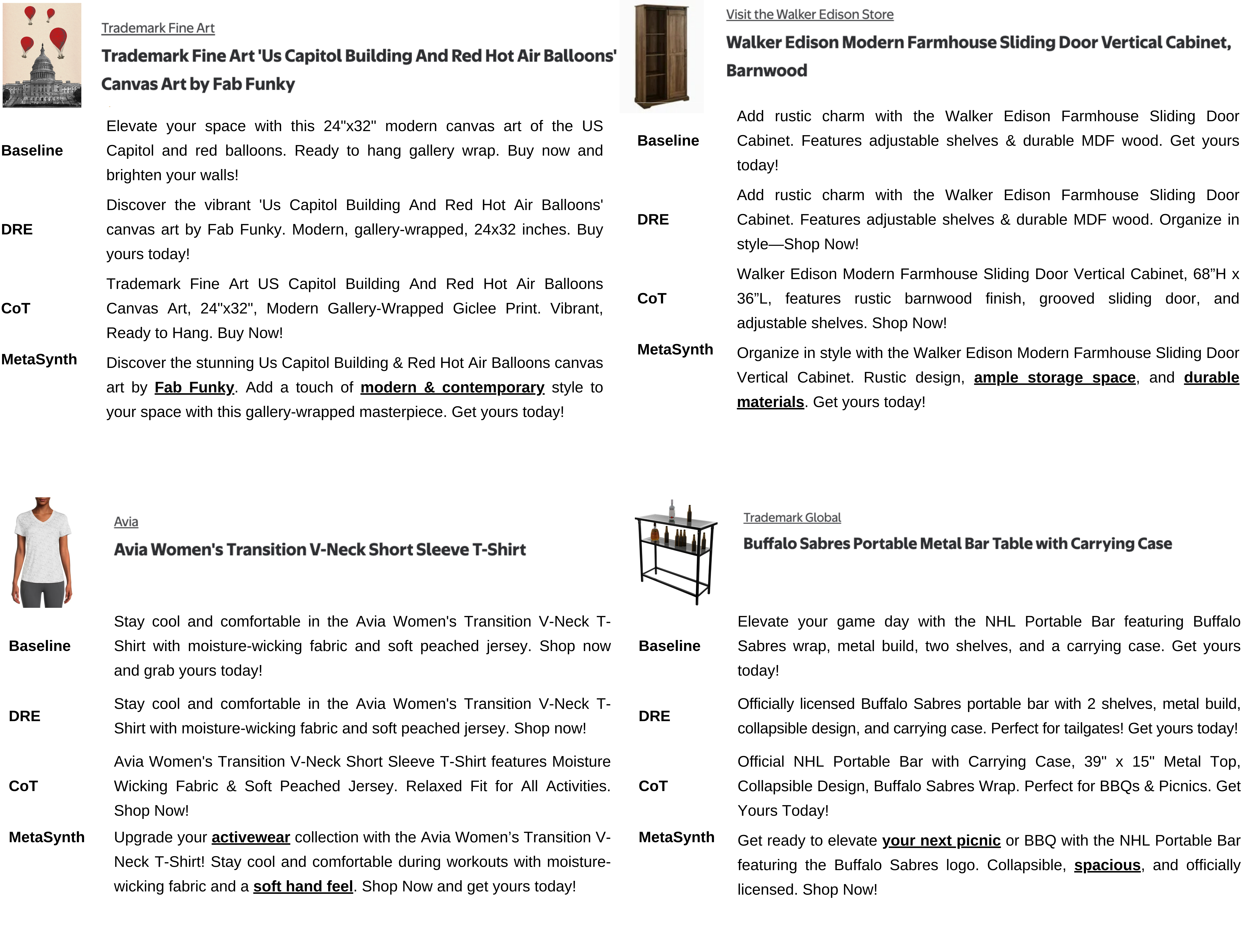}
    \caption{\small Case studies comparing MetaSynth method with three other studies (baseline, DRE, COT), in which MetaSynth outcome ranked first among others in the search engine. Some influential words that might be affected the ranking are demonstrated with bold font. Across four verticals of home décor, furniture, apparel, and tailgating/barware, MetaSynth replaces specification lists with fluent, benefit‑led phrasing while preserving key retrieval tokens. This selective emphasis improves readability and broadens intent matching without keyword stuffing, aligning with the relevance–readability–compliance objectives of the method.}
    \label{fig:case-study}
\end{figure*}

\subsection{Ablation Study}
\begin{table}[htbp!]
\centering
\caption{Ablation Studies}
\begin{tabular}{lccc}
\toprule
\textbf{Approach} & \textbf{Average rank}   & \textbf{NDCG} &\textbf{ MRR} \\ 
\midrule
MetaSynth wo RAG         & 2.4830& 0.6245& 0.5018\\ 
Meta Synth wo Evaluation & 1.9769       & 0.7267   & 0.6353 \\ 
\textbf{MetaSynth}                 & \textbf{1.6416}& \textbf{0.7835}& \textbf{0.7204}\\ 
\bottomrule
\label{Table:Ablation}
\end{tabular}
\end{table}

To systematically analyze the contribution of individual components within our pipeline, we conducted ablation studies across 3 variants i) MetaSynth without  Library Construction Module ii)  MetaSynth without Library Construction and Evaluation modules iii) Complete MetaSynth pipeline. For this analysis, we compared the performance of each these variants with benchmark models on our Proprietary dataset.

The results summarized in (see Table \ref{Table:Ablation} ) show that exclusion of Library Construction Module leads to a substantial degradation in performance there is a 33\% drop in average rank compared to the complete Meta Synth framework  which  highlights  the role of high quality  exemplar titles and descriptions. Additionally, when we remove the  Evaluation and Feedback module, there is   again a considerable  decline in all metrics indicating  that the loop of evaluation, feedback consolidation and regeneration  is essential for good performance. In ranking effectiveness, MetaSynth yields a 25.5\% gain in MRR and a 25.4\% gain in NDCG compared to MetaSynth w/o RAG, while achieving a further 13.4\% (MRR) and 7.8\% (NDCG) improvement over MetaSynth w/o Evaluation.
Notably,  the observed performance drop is more pronounced when Library Module is omitted compared to ii) W/o Evaluation suggesting that  Library Module exerts a greater influence on the performance, Overall, having all three  components  surpasses the other methods demonstrating the  significance of each of the components. In particular, evaluation modeuls enhances ranking precision, while Library Construction Module enriches the VectorDB enabling Meta Synth to perform the best in all  offline evaluation metrics.

\subsection{Human-Annotated Evaluation}
To productionize the generated meta titles and descriptions and expose them to live traffic, we first ran a human annotation study. We were required to satisfy predetermined minimum quality thresholds set by the quality-control team dictated by our product launch protocols. These proprietary requirements were divided into hard and soft constraints. Results are shown in table \ref{tab:manual_eval},

\begin{table}[h]
\centering
\caption{Human evaluation pass rates for generated meta-title and description by MetaSynth.}
\label{tab:manual_eval}
\begin{tabular}{lcc}
\toprule
 & \textbf{Soft Criteria} & \textbf{Hard Criteria} \\
\midrule
Meta-title  & 98.2\% & 100.0\% \\
Meta-descriptions  & 96.6\% & 100.0\% \\
\bottomrule
\end{tabular}
\end{table}
As shown in the table, all the generations of MetaSynth pass the given hard criteria and obtaining near 100\% pass rate for the soft criteria, which confirms MetSynth as a production grade option for generating meta-title and descriptions.

\subsection{A/B Test Results}
To evaluate the impact of meta-titles and snippets on user engagement, we conducted 4-weeks long A/B test to compare a propreitary  control search engine meta generator with the ones generated through MetaSynth. In the process of generating and evaluating MetaSynth titles and descriptions, we used OpenAI's \textit{gpt-3.5-turbo} model for each LLM evaluator, and refinement agent. For titles, there were 3 evaluator agents to evaluate relevance, uniqueness, and specific details. For description, there were 4 evaluator agents to evaluate presence of selling points, call to action, duplicates, and relevance. Given the feedback from each evaluator agent, a separate refinement agent would regenerate the title/description and the refined title/description would go through another round of evaluation. The evaluation and refinement continued in a loop until at least 99\% satisfactory title/description across all evaluators. With the above structure we could extend the meta title and generation to $O(\text{millions})$ of gnerations for the AB test.

As shown in Table \ref{abtest:table}, MetaSynth leads to a +10.26\% improvement in clicks and 7.51\% in CTR when compared to the control model.
These statistically significant lifts confirm that MetaSynth’s offline improvements translate directly into real-world user engagement. Importantly, the gains substantially outweigh the additional inference costs introduced by the multi-agent pipeline, demonstrating that MetaSynth is both effective and scalable for large catalogs.
\begin{table}[h]
\centering
\begin{tabular}{lcc}
\toprule
\textbf{Metric} & \textbf{Lift}  \\
\midrule
CTR   & +10.26\% \\
Overall Clicks  & +7.51\% \\
\bottomrule
\end{tabular}
\caption{A/B test performance}
\label{abtest:table}
\end{table}

\subsection{Case Studies}
Figure \ref{fig:case-study} reviews meta descriptions for four products; the top‑left case (“US Capitol Building and Red Balloons”) exemplifies the pattern. Relative to Baseline, DRE, and CoT, MetaSynth achieves greater lexical economy and naturalness: it foregrounds the core entity and creator (“…by Fab Funky”), compresses qualifiers (“modern \& contemporary style”), and uses one semantically dense phrase (“gallery‑wrapped masterpiece”) instead of loosely coupled attribute lists. Baseline omits the brand/creator and relies on generic phrasing (“brighten your walls”); DRE reduces to slot‑filled keywords (“Modern, gallery‑wrapped, 24x32 inches”); and CoT over‑enumerates proper nouns and media terms (“Trademark Fine Art … Canvas Art … Giclée Print”), resembling inventory metadata rather than user‑facing copy. This shift from enumeration to fluent phrasing improves readability while retaining retrieval‑salient tokens (e.g., “gallery‑wrapped,” style cues), aligning with relevance, readability, and technical‑compliance goals. MetaSynth also balances intent coverage with precision: “modern \& contemporary style” broadens matchability without keyword stuffing, and naming the creator “Fab Funky” adds a trusted cue that helps disambiguate entity searches.

Additionally, we compare the four variants for the ``Walker Edison Farmhouse Sliding Door Cabinet" product description (See Figure \ref{fig:case-study}, top-right case). Relative to the Baseline, DRE, and CoT texts, the MetaSynth description demonstrates superior lexical efficiency: it leads with action-oriented framing (``Organize in style"), emphasizes core value propositions (ample storage space, durable materials), and maintains brand recognition while avoiding specification overload. In contrast, Baseline and DRE rely on identical phrasing, while CoT overwhelms with technical specifications (68"H x 36"L, ``grooved sliding door") that read as catalog entries rather than persuasive copy. MetaSynth's selective emphasis highlights functionality and longevity without dimensional clutter, while ``Organize in style" efficiently combines utility with aesthetic appeal. This creates broader search matching than CoT's specification-heavy approach while maintaining more substance than Baseline's generic ``rustic charm" language, improving both readability and conversion potential.

For the Avia Women’s Transition V‑Neck T‑Shirt’’ (Figure \ref{fig:case-study}, bottom-left), MetaSynth shows stronger market positioning than Baseline, DRE, and CoT. It opens with aspirational framing (Upgrade your activewear collection''), spotlights category context (activewear) and tactile appeal (soft hand feel), and embeds technical benefits in engaging copy. Baseline and DRE rely on generic comfort claims; CoT lists features (Moisture Wicking Fabric \& Soft Peached Jersey. Relaxed Fit''), prioritizing specifications over lifestyle. MetaSynth’s selective bolding clarifies category intent and translates peached jersey’’ into consumer‑friendly sensory language. By casting the item as an upgrade rather than a necessity, it broadens intent matching and strengthens purchase motivation, improving category relevance and conversion potential.

Finally, we compare the four variants for the ``NHL Portable Bar featuring Buffalo Sabres" product description (See Figure \ref{fig:case-study}, bottom-right). Relative to the Baseline, DRE, and CoT texts, the MetaSynth description demonstrates superior contextual targeting: it leads with experiential framing (``Get ready to elevate your next picnic"), emphasizes specific use scenarios (your next picnic) and key functional benefits (spacious), while streamlining technical details into essential selling points. In contrast, Baseline opens with generic ``game day" positioning, DRE focuses heavily on structural specifications, while CoT overwhelms with dimensional data (39" x 15") and feature lists that prioritize inventory details over lifestyle appeal. MetaSynth's selective bolding of your next picnic creates direct personal relevance and expands beyond traditional sports contexts, while spacious translates technical shelf configurations into practical consumer benefit language. The phrase ``Get ready to elevate your next picnic" positions the product within broader outdoor entertainment rather than limiting to sports events, creating wider intent matching than Baseline's ``game day" restriction or CoT's specification-heavy approach, improving both contextual relevance and purchase motivation.

\section{Conclusion}

We presented \textbf{MetaSynth}, a multi-agent retrieval-augmented generation framework for optimizing metadata in search and recommendation settings. Unlike prior template-based, prompt-only, or standard RAG approaches, MetaSynth directly leverages implicit signals from observable outcomes, treating top-ranked results as weak supervision for learning style and content preferences. Our design integrates three key components: an exemplar library, a constrained generator conditioned on product content and exemplars, and an evaluator--generator refinement loop that enforces relevance, promotional strength, and compliance.  

Extensive experiments across both proprietary e-commerce data and the Amazon Reviews corpus demonstrate consistent gains in NDCG, MRR, and ranking quality compared to strong baselines. Large-scale online A/B tests further confirm the practical impact, yielding \textbf{+10.26\% CTR} and \textbf{+7.51\% clicks}. These results highlight MetaSynth’s ability to bridge the gap between black-box ranking systems and generative optimization, offering a reproducible methodology that is both effective and deployable.

Beyond metadata generation, this work contributes a broader paradigm for learning from implicit feedback in black-box environments. By showing how multi-agent generation can integrate weak supervision, retrieval, and iterative critique, we open new directions for applying similar techniques to recommendation, personalization, and ranking-adjacent tasks. Future research may extend this framework to richer modalities (e.g., images, video), integrate counterfactual debiasing of implicit signals, and explore theoretical guaranties for convergence under noisy supervision.

\bibliographystyle{abbrv}
\bibliography{references}

\end{document}